\documentclass{elsart}

\usepackage{adnd}
\usepackage{longtable}

\usepackage{mathptmx}

\usepackage{amsmath}

\usepackage{amssymb,amstext}
\usepackage[pdftex,
        letterpaper=true,
        hyperindex=true,
        breaklinks=true,
        colorlinks=false,
        citecolor=blue,
        pdftitle={},
        pdfauthor={}]
{hyperref}

\usepackage{graphicx}

\begin{document}

\begin{frontmatter}

\journal{Atomic Data and Nuclear Data Tables}

\copyrightholder{Elsevier Science}

\runtitle{Cerium}
\runauthor{Ginepro}


\title{Discovery of the Cerium Isotopes}


\author{J.Q. Ginepro},
\author{J.~Snyder},
\and
\author{M.~Thoennessen\corauthref{cor}},\corauth[cor]{Corresponding author.}\ead{thoennessen@nscl.msu.edu}

\address{National Superconducting Cyclotron Laboratory, and \\ Department of Physics and Astronomy, Michigan State University, \\East Lansing, MI 48824, USA}

\date{15.11.2008} 
\begin{abstract}
The discovery of the 35 cerium isotopes discovered up to date is discussed. For each isotope a brief summary of the first refereed publication, including the production and identification method, is presented.
\end{abstract}

\date{24.12.2008} 

\begin{abstract}
The discovery of the 35 cerium isotopes presently known is discussed. Criteria for the discovery of isotopes are suggested and for each isotope a brief summary of the first refereed publication, including the production and identification method, is presented.
\end{abstract}

\end{frontmatter}





\newpage
\tableofcontents
\listofDtables

\vskip5pc

\section{Introduction}\label{s:intro}

In recent years, the knowledge about individual isotopes has been increasing rapidly. The most recent table of isotopes consists of two volume of over 3000 pages and it has been already 12 years since the last edition has been published \cite{Fir96} with an updated version in 1999 \cite{Fir99}. In this wealth of information, the history of the discovery of many of the isotopes can get lost.

\begin{figure}
	\centering
	\includegraphics[width=10cm]{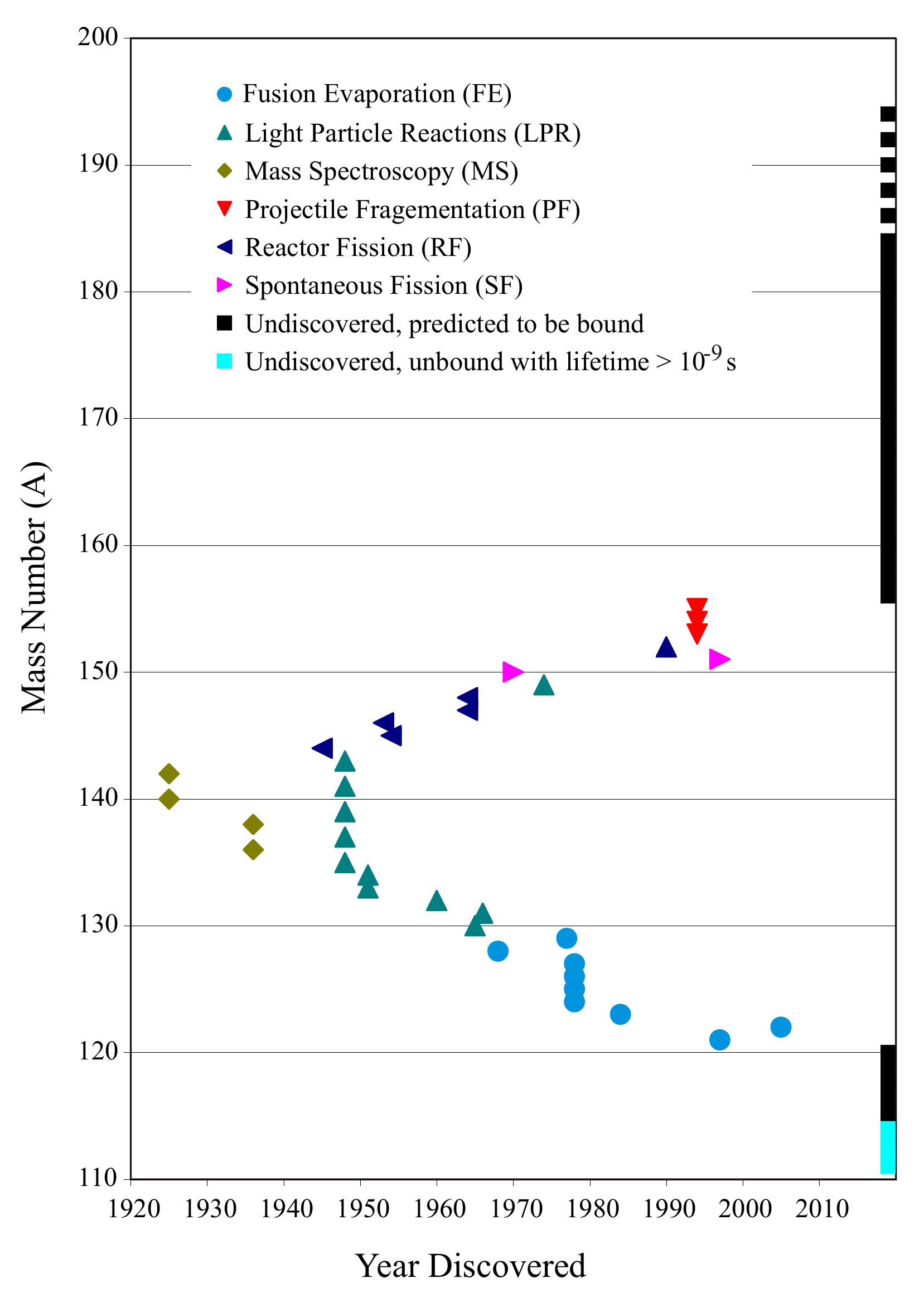}
	\caption{Cerium isotopes as a function of time they were discovered. The solid black squares on the right hand side of the plot are isotopes predicted to be bound by the HFB-14 model.  On the proton-rich side the light (blue) squares correspond to unbound isotopes predicted to have lifetimes larger than $\sim 10^{-9}$ s.}
	\label{f:year}
\end{figure}

In contrast to the discovery of a new element \cite{Wap91,Bar92} the first observation of a new isotope is not as well defined. The criteria for the discovery of an element were established in 1991: ``Discovery of a chemical element is the experimental demonstration, beyond reasonable doubt, of the existence of a nuclide with an atomic number Z not identified before, exiting for at least 10$^{-14}$s.'' \cite{Wap91}. For isotopes, even the question of existence does not have an obvious answer \cite{Tho04}.

In the process of summarizing the discovery of isotopes we decided on two main guidelines. (1) In order to qualify as a discovery the isotope had to be cleanly identified, either by decay curves and relationships to other known isotopes, particle or $\gamma$-ray spectra, or unique mass and Z identification. (2) The discovery had to be reported in a refereed journal. Publications in conference proceedings and unpublished reports or theses were not considered as a discovery.

The first criterion excludes measurements of half-lives of a given element without mass identification. This affects mostly isotopes first observed in fission where decay curves of chemically separated elements were measured without the capability to determine their mass. Also the four-parameter measurements (see for example \cite{Wat70,Joh70}) were in general not considered because the mass identification was only $\pm 1$ mass units.

The second criterion affects especially the isotopes studied within the Manhattan Project. Although an overview of the results was published in 1946 \cite{PPR46} most of the papers were only published in the Plutonium Project Records of the Manhattan Project Technical Series, Volume 9A {\it Radiochemistry and the Fission Products} in three books by Wiley in 1951 \cite{PPR51}. We considered this first unclassified publication to be equivalent to a refereed paper.

\section{Discovery of $^{121-155}$Ce}

As a first element we present the discovery of cerium isotopes. Up to now 35 cerium isotopes from A = 121 $-$ 155 have been discovered. There are 4 stable, 17 proton-rich and 14 neutron-rich isotopes. According to the latest HFB model \cite{Gor07} $^{115}$Ce is predicted to be the lightest and $^{194}$Ce the heaviest particle-bound cerium isotope. In addition about four proton-rich isotopes are estimated to exist long enough ($>10^{-9}$s) to be observed directly \cite{Tho04}. Thus less than half of all possible cerium isotopes have been observed.

Figure \ref{f:year} summarizes the year of first discovery for all cerium isotopes identified by the production method. The stable isotopes were identified by mass spectroscopy (MS). Isotopes close to stability and more proton-rich isotopes were first produced by reactions with light particles (LPR) while more proton-rich nuclei were all discovered in heavy-ion fusion-evaporation reactions (FE). Heavy ions are all nuclei with an atomic mass larger than A=4 \cite{Gru77}. Light particles also include neutrons produced by accelerators (see $^{149}$Ce). The neutron-rich cerium isotopes were discovered in fission following irradiations in reactors (RF), or spontaneous fission (SF), and more recently projectile fragmentation or projectile fission (PF).

In the following the discovery of each cerium isotope is discussed in detail.
When a discovery included a half-life measurement the measured value is compared to the currently adapted value taken from the NUBASE evaluation \cite{Aud03} which is based on the ENSDF database \cite{ENS08}. When appropriate, references to conference proceedings, internal reports, and theses are included.

\subsection*{$^{121}$Ce}\vspace{-.85cm}
$^{121}$Ce was discovered in 1997 at the Institute of Modern Physics in Lanzhou, China, by Li Zhankui {\it et al.} reported in the paper {\it $\beta$-delayed proton precursor $^{121}$Ce} \cite{Zha97}. The new isotope was produced during the bombardments of a $^{92}$Mo target with a 171-MeV $^{32}$S beam. The residues of the fusion-evaporation reaction $^{92}$Mo($^{32}$S,3n) were transported to a counting station using a helium-jet recoil fast-moving tape-transport system. Coincidence measurements of $\beta$-delayed protons and $\gamma$ rays were performed. ``Thus, the 186-keV $\gamma$-ray transition in the daughter nucleus $^{120}$Ba, was here used to identify its $\beta$-delayed proton precursor $^{121}$Ce.'' The measured half-life of $^{121}$Ce was 1.1(1)s, which is currently the only half-life measurement of this isotope.

\subsection*{$^{122}$Ce}\vspace{-0.85cm}
The {\it First observation of very neutron-deficient $^{122}$Ce} was reported in 2005 by Smith {\it et al.} \cite{Smi05}. A 260 MeV $^{64}$Zn from the ATLAS facility at Argonne National Laboratory was used to produce $^{122}$Ce in the fusion-evaporation reaction $^{64}$Zn($^{64}$Zn,$\alpha$2n). Gamma-rays were measured with Gammasphere in coincidence with neutrons and charged particles detected in the Washington University Neutron Shell and Microball, respectively. ``The excited states have been unambiguously assigned to $^{122}$Ce by detecting gamma-ray transitions in coincidence with the evaporated charged particles and neutrons.'' $^{122}$Ce is currently the last cerium isotope being discovered.

\subsection*{$^{123}$Ce}\vspace{-0.85cm}
In the 1984 article {\it Beta-Delayed Proton Emission Observed in New Lanthanide Isotopes} Nitschke {\it et al.} reported the first observation of $^{123}$Ce \cite{Nit84}. It was produced in the fusion-evaporation reaction $^{92}$Mo($^{36}$Ar,$\alpha$n) with a 196 MeV $^{36}$Ar beam from the SuperHILAC at Lawrence Berkeley National Laboratory. Beta-delayed protons and characteristic X-rays were measured in coincidence at the on-line isotope separator OASIS. ``The energies of the two-x-ray lines are in excellent agreement with the literature values for La K$_\alpha$- and K$_\beta$-radiation, which uniquely identifies the new isotope as $^{123}$Ce.'' The extracted half-life was 3.8(2)s which is presently the only published value for $^{123}$Ce.

\subsection*{$^{124-127}$Ce}\vspace{-0.85cm}
The discovery of the isotope $^{124}$Ce, $^{125}$Ce, $^{126}$Ce, and $^{127}$Ce was first presented in the paper {\it  New Neutron-Deficient Isotopes of Lanthanum and Cerium} by Bogdanov {\it et al.} \cite{Bog78}. A 190 MeV $^{36}$S beam accelerated by the U-300 heavy-ion cyclotron of the Joint Institute for Nuclear Research (JINR) facility at Dubna, bombarded targets of $^{96}$Ru and $^{98}$Ru. The fusion-evaporation residues were mass separated with the on-line BEMS-2 facility and their X-ray and $\beta$ emission was detected with e Ge(Li) spectrometer and a plastic counter, respectively. Half-lives were determined from the X-ray decay curves. ``Seven isotopes $^{123-125}$La and $^{124-127}$Ce have been first observed and their half-lives and low-energy $\gamma$-ray data are reported.'' The half-life of 6(2)s for $^{124}$Ce is included in the average for the adapted value of 9.1(12)s. The measured half-lives for $^{125}$Ce and $^{126}$Ce of 11(4)s and 50(6)s are included in the presently adapted weighted average of 9.3(3)s and 51.0(3)s, respectively. The half-life of 32(4)s for $^{127}$Ce agrees well with the current value of 31(2)s.

\subsection*{$^{128}$Ce}\vspace{-0.85cm}
The first experimental evidence of $^{128}$Ce was discussed in the paper {\it Collective Levels in Light Even Ce Isotopes} by Ward {\it et al.} published in 1968 \cite{War68}. Beams of 80 MeV $^{16}$O and 90 MeV $^{20}$Ne were produced by the heavy-ion accelerator HILAC at Berkeley to form $^{128}$Ce in the fusion-evaporation reactions $^{116}$Sn($^{16}$O,4n) and $^{112}$Cd($^{20}$Ne,4n), respectively. Four $\gamma$-ray transitions were measured with two lithium-drifted germanium counters and a level scheme was proposed. The paper does not mention that it represents the first identification of $^{128}$Ce. It can only be speculated that the authors were aware of a contribution to a conference proceeding reporting half-life measurement of $^{128}$Ce \cite{Arl69a}.

\subsection*{$^{129}$Ce}\vspace{-0.85cm}
The first identification of $^{129}$Ce was reported in 1977 by Gizon {\it et al.} \cite{Giz77} in the paper {\it The h$_{11/2}$ and g$_{7/2}$ band structure in $^{131}$Ce and $^{129}$Ce}. The isotope was produced in the fusion-evaporation reaction $^{116}$Sn($^{16}$O,3n) with an $^{16}$O beam from the 88'' sector-focussed cyclotron at Berkeley. $^{129}$Ce was identified by $\gamma \gamma$ coincidence measurements of excitation functions. Two prior reports of half-life measurements mentioned in the paper ``Before this work, no information was available on the energies and spin assignments of levels in $^{131}$Ce and $^{129}$Ce; only their half-lives had been determined'' were not published in refereed journals. In addition, the measured half-lives of approximately 13 min \cite{Lav63} and 3.5(5) min \cite{Arl69b} are in disagreement. A later measurement of 3.5(3) min \cite{Alk93} was not included in the ENSDF data evaluation.

\subsection*{$^{130}$Ce}\vspace{-0.85cm}
Gerschel {\it et al.} reported the first observation of $^{130}$Ce in the 1965 paper {\it Nuclear Spectroscopy of Some Neutron-Deficient Even Barium Isotopes} \cite{Ger65}. The Cerium activities were produced by 100-150 MeV proton bombardment of lanthanum in (p,xn) reactions at Orsay, France. The assignment was based on $\gamma$-scintillation spectra of chemically separated samples. The paper concentrates on the properties of $^{130}$Ba and $^{132}$Ba: ``...experiments have been performed at Orsay on the decay of new cerium activities...'' referring to a 1964 thesis. The extracted half-life of 30m is close to the accepted value of 22.9(5)m. It was later pointed out \cite{Nor66} that an earlier observation of a 30m half-life of a cerium isotope was most likely $^{130}$Ce \cite{War60} (see $^{131}$Ce).

\subsection*{$^{131}$Ce}\vspace{-0.85cm}
The 1966 paper {\it On the Decay of $^{131}$Ce} by Norris {\it et al.} reported the first conclusive assignment of $^{131}$Ce produced at Brookhaven National Laboratory \cite{Nor66}. The isotope was generated with 40 MeV $^4$He and 20 MeV $^3$He beams from the Brookhaven 150 cm cyclotron in the reactions $^{130}$Ba($^{4}$He,3n) and $^{130}$Ba($^{3}$He,2n), respectively. Gamma-ray spectra of chemical separated samples were measured with multi-channel scintillation spectrometers. ``A 10 min activity in chemically separated cerium was identified as $^{131}$Ce by timed separation of its $^{131}$La dn $^{131}$Ba descendants.'' The direct cerium decay measurement resulted in a half-life of 10.5(6) min while the timed lanthanum separation experiment yielded 9.5(3)m. Although the abstract of the paper quotes 10m without uncertainties, only the former value is included in the weighted average of the currently accepted value of 10.2(3)m. In an earlier measurement a half-life of 30 min was speculated to correspond to $^{131}$Ce \cite{War60}. However, the authors made no assignment: ``An activity of 30-minutes half-life and 4.2 MeV maximum positron energy, which may be due to $^{131}$Ce, was also observed'' and ``The identity of the 30-minute activity is open to question''. Norris {\it et al.} pointed out that this activity was most likely due to $^{130}$Ce.

\subsection*{$^{132}$Ce}\vspace{-0.85cm}
In the paper {\it New Neutron-Deficient Isotope of Cerium} where Ware and Wiig speculated about the observation of $^{131}$Ce, they reported the discovery of $^{132}$Ce in \cite{War60}. Cerous oxalate was bombarded in the 130-inch synchrocyclotron of the University of Rochester by 240 MeV protons. The assignment was made based on positron radiation from a chemically separated sample and the known properties of the $^{132}$La daughter. ``This result was consistent with the work described above with the cerium fraction and is taken as a proof of the generic relationship between a 4.2-hour activity and $^{132}$La, its daughter. $^{132}$Ce is the only possible isotope.'' The half-life of 4.2(2)h is close to the presently accepted value of 3.51(11)h.

\subsection*{$^{133,134}$Ce}\vspace{-0.85cm}
Stover reported the discovery of the two new isotopes $^{133,134}$Ce in {\it New Neutron-Deficient Radioactive Isotopes of the Light Rare-Earth Region} in 1951 at the University of California at Berkeley \cite{Sto51}. Lanthanum oxide was bombarded with 60 to 80 MeV protons and the isotopes $^{133}$Ce and $^{134}$Ce were produced in the reactions $^{139}$La(p,7n) and $^{139}$La(p,6n), respectively. They were identified by absorption curve and magnetic counter methods. ``The 4.0-hr $^{133}$La, which has been run on the mass spectrograph, was shown to grow in only during the decay of the 6.3-hr activity which must then be $^{133}$Ce'' and ``That the $^{134}$La daughter [$^{134}$Ce] emitted the positions was verified by following the decay of a separated $^{134}$La asmple on the magnetic counter.'' The observed half-life for $^{133}$Ce of 6.3(1)h corresponds most likely to the decay of the 9/2$^-$ isomer which has an adapted half-life of 4.9(4)h. The half-life of 72.0(5)h extracted for $^{134}$Ce compares to an adapted value of 75.9(9)h.

\subsection*{$^{135}$Ce}\vspace{-0.85cm}
$^{135}$Ce was discovered by Chubbuck and Perlman at Berkeley in the 1948 published paper {\it Neutron Deficient Isotopes of Cerium and Lanthanum} \cite{Chu48}. 60 MeV deuterons bombarded a La$_2$O$_3$ target to produce $^{135}$Ce in the reaction $^{139}$La(d,6n). It was created by bombarding La$_2$O$_3$ with 60 MeV deuterons. Decay curves were measured with a beta-ray spectrometer and the assignment was based on the absence of the newly observed activity at lower bombarding energies: ``The activity is assigned to $^{135}$Ce because it had not appeared in bombardments at lower energies, and 19.5-hour $^{135}$La could be shown to grow into the cerium fraction at a rate corresponding to a half-life for its parent of about 16h.'' The half-life of ``about 16h'' compares well with the adapted value of 17.7(3)h.

\subsection*{$^{136}$Ce}\vspace{-0.85cm}
The stable isotope $^{136}$Ce was discovered by Dempster from the University of Chicago in his 1936 paper ``The Isotopic Constitution of Barium and Cerium'' \cite{Dem36}. He took photographs of mass spectra of cerium ions formed in a high frequency spark using a pure cerium oxide sample. ``With cerium two faint isotopes at 136 and 138 have also been observed, in addition to the two strong ones at 140 and 142.''

\subsection*{$^{137}$Ce}\vspace{-0.85cm}
The discovery of $^{137}$Ce was reported by Chubbuck and Perlman in the same publication {\it Neutron Deficient Isotopes of Cerium and Lanthanum} \cite{Chu48} where they presented the discovery of $^{135}$Ce. 20 MeV deuterons from the 60-inch cyclotron and 40 MeV deuterons from the 184-inch cyclotron at Berkeley were used to produce $^{137}$Ce in the reaction $^{139}$La(d,4n). The lanthanum fraction was separated and $\beta$- and X-ray spectra were measured. The observation from the 20 MeV data: ``A sample of the chemically separated cerium resolved into two components: 140-day $^{139}$Ce and a 36-hour activity assigned below to $^{137}$Ce.'' were confirmed with the 40 MeV data: ``...and the most likely assignment for the 36-hour activity is $^{137}$Ce''. The 36 hr half-life assigned to $^{137}$Ce, corresponds to the decay of the 11/2$^-$ isomeric state at 254 keV and agrees with the adapted value of 34.4(3)h.

\subsection*{$^{138}$Ce}\vspace{-0.85cm}
The stable isotope $^{138}$Ce was discovered by Dempster from the University of Chicago in the same 1936 paper {\it The Isotopic Constitution of Barium and Cerium} \cite{Dem36} where he also observed $^{136}$Ce. He took photographs of mass spectra of cerium ions formed in a high frequency spark using a pure cerium oxide sample. ``With cerium two faint isotopes at 136 and 138 have also been observed, in addition to the two strong ones at 140 and 142.''

\subsection*{$^{139}$Ce}\vspace{-0.85cm}
$^{139}$Ce was discovered at Ohio State University in 1948 by Pool and Krisberg: {\it Radioactive Cerium and Praseodymium} \cite{Poo48}. 10 MeV deuterons from the 42-inch cyclotron produced $^{139}$Ce in the reaction $^{139}$La(d,2n). A decay curve with a halflife of 140(1)d was observed in the $\gamma$- and $\beta$-decay. The assignment was based on the observation of $K_\alpha$ x-rays from lanthanum: ``The 140-day half-life is therefore assigned to $^{139}$Ce.'' This half-life (140(1)d) is consistent with the adapted value of 137.641(20)d. It is curious to note that the authors do not mention an earlier paper by the same first author \cite{Poo38} where they reported a 2.1 min positron activity ``...the carrier of the 2.1-min. positron activity is very probably $^{139}$Ce.'' Although it could be possible that they observed the isomeric state of $^{139}$Ce, the half-life of this state (54.8(10)s) is significantly different.

\subsection*{$^{140}$Ce}\vspace{-0.85cm}
The stable isotope $^{140}$Ce was discovered together with $^{142}$Ce in 1925 by Aston at the Cavendish Laboratory in Cambridge, England during mass spectroscopy experiments: {\it The Mass-Spectra of Chemical Elements - Part VI. Accelerated Anode Rays Continued} \cite{Ast25}. Two cerium isotopes were identified using cerium bromide with a strong mass line at 140 and a weaker one at 142: ``Cerium may therefore be taken as a complex element with mass numbers 140 and 142.'' With the isotopic identification this paper represents the first discovery of any cerium isotope.

\subsection*{$^{141}$Ce}\vspace{-0.85cm}
$^{141}$Ce was discovered at Ohio State University in 1948 by Pool and Krisberg in the same paper where they reported the discovery of $^{139}$Ce: {\it Radioactive Cerium and Praseodymium} \cite{Poo48}. $^{141}$Ce was created by the bombardments of barium with 20 MeV $\alpha$ particles ($^{138}$Ba($\alpha$,n)) and cerium with 10 MeV deuterons ($^{140}$Ce(d,p)) and fast neutrons ($^{142}$Ce(n,2n)). Although this activity could also have been assigned to $^{140}$Ce, the absence of the same activity from deuteron bombardment of lanthanum, eliminated the possible assignment to $^{140}$Ce. ``The 30.6-day activity is therefore assigned $^{141}$Ce and decays to stable $^{141}$Pr...'' The reported half-life of 30.6(7)d is consistent with the adapted half-life of 32.508(13)d.

\subsection*{$^{142}$Ce}\vspace{-0.85cm}
The stable isotope $^{142}$Ce was discovered together with $^{140}$Ce in 1925 by Aston at the Cavendish Lab in Cambridge, England, during mass spectroscopy experiments: {\it The Mass-Spectra of Chemical Elements - Part VI. Accelerated Anode Rays Continued} \cite{Ast25}. Two cerium isotopes were identified using cerium bromide with a strong mass line at 140 and a weaker one at 142: ``Cerium may therefore be taken as a complex element with mass numbers 140 and 142.'' With the isotopic identification this paper represents the first discovery of any cerium isotope.

\subsection*{$^{143}$Ce}\vspace{-0.85cm}
The paper {\it Radioactive Cerium and Praseodymium} by Pool and Krisberg from 1948 reporting the discovery of $^{139}$Ce and $^{141}$Ce also presented the first observation of $^{143}$Ce at Ohio State University \cite{Poo48}. The $^{143}$Ce activity was observed by subtracting the 30.6d activity of $^{141}$Ce from a sample of cerium that had been bombarded with 10 MeV deuterons ($^{142}$Ce(d,p)) and neutrons ($^{142}$Ce(n,$\gamma$)). The paper states: ``Since the 1.4-day activity in cerium could be produced by a deuteron bombardment of cerium but not by an alpha-particle bombardment of barium, it is assigned to $^{143}$Ce...'' The observed activity of 1.4(1)d agrees with the adapted value of 33.039(6)h.

\subsection*{$^{144}$Ce}\vspace{-0.85cm}
$^{144}$Ce was identified Burgus {\it et al.} as part of the Plutonium Project {\it Characteristics of the 275d $^{144}$Ce} at Argonne National Laboratory. The date of the last internal report contributing to this paper was 1945 \cite{Bur45}. Two decay curves were measured following chemical separation of fission products from thermal neutron radiation of $^{235}$U. The average half-life was 275d which is close to the adapted value of 284.91(5)d. First evidence for this long-lived cerium activity was reported by Hahn and Strassman \cite{Hah40}, however, they could not identify the isotope: ``Wir glauben, zwei Cer-Isotope, eines von etwas 20 Tagen und eines von 200 Tagen, letzteres in sehr roher Sch\"atzung, nachgewiesen zu haben.'' (We believe that we observed two cerium isotopes, one with 20 days and one with 200 days, the latter being a very rough estimate.) Burgus {\it et al.} state: ``The long half-life and the weak $\beta$ energy of the cerium parent suggest that the mass assignment is 144, and this has been by Lewis and Hayden by mass spectrometry.'' The Lewis and Hayden refers to an internal report \cite{Lew45} which is not included in the published papers, thus we credit Burgus {\it et al.} with the identification of $^{144}$Ce.

\subsection*{$^{145}$Ce}\vspace{-0.85cm}
$^{145}$Ce was discovered by Markowitz {\it et al.} in 1954 at Brookhaven National Laboratory: {\it A New 3.0min Ce Fission Product and its 5.95-hr Pr Daughter}. Uranyl nitrate was irradiated with neutrons in the Brookhaven pile and the cerium and praseodymium was chemically separated. A 3.0 min activity was assigned to $^{145}$Ce by correlating it the 6 hour activity of $^{145}$Pr: ``The observed decrease in 6-hr Pr activity with each successive sample shows that it is the daughter of a 3.0-min Ce.'' The measured activity of 3.0(1)m is included in the weighted average of the adapted value of 3.01(6)m. A previously reported 1.8 hour activity attributed to $^{145}$Ce \cite{Bal51} was shown to be incorrect \cite{Car53}.

\subsection*{$^{146}$Ce}\vspace{-0.85cm}
$^{146}$Ce was identified in 1953 by Caretto and Katcoff, at Brookhaven National Laboratory in the paper entitles {\it Short-Lived Cerium Isotopes from Uranium Fission} \cite{Car53}. $^{146}$Ce was produced from irradiation of uranyl nitrate and decay curves were measured with an end-window Geiger tube. The measured half-life of 13.9(6) min is included in the weighted average of 13.52(13)m. Caretto and Katcoff did not consider their measurement a new discovery: ``$^{146}$Ce and $^{146}$Pr were briefly reinvestigated.'' They quote an unpublished report of the Fission Studies by Schuman \cite{Sch45} reporting a half-life of 14.6(8) min which is also included in the adapted weighted average. They also mention the measurement by H. G\"otte \cite{Got46} who observed an 11 min half-life of cerium but did not make a mass assignment: ``Mit Hilfe der geschilderten Abscheidungsmethode wurde[n] ein neues Cer-Isotop von 11m Halbwertszeit ... aufgefunden.'' (With help of the described separation method a new cerium isotope with a half-lilfe of 11m was found.)

\subsection*{$^{147,148}$Ce}\vspace{-0.85cm}
The two cerium isotopes $^{147}$Ce and $^{148}$Ce wer discovered by Hofman and Daniels at Los Alamos Scientific Laboratory in 1964: {\it Some Short-Lived Isotopes of Cerium and Praseodymium} \cite{Hof64}. Uranyl nitate was irradiated by the Los Alamos Water Boiler Reactor. ``The cerium half-lives were determined by measuring the relative amounts of the praseodymium daughter activities in samples milked from fission product cerium at suitable intervals.'' The $^{145}$Ce and $^{146}$Ce were 65(6) s and 43(10) s, respectively, in reasonable agreement with the adapted values of 56.4(10) s and 56(1) s.

\subsection*{$^{149}$Ce}\vspace{-0.85cm}
$^{149}$Ce was tentatively identified in 1974 by Aronsson {\it et al.} from Chalmers University of Technology in G\"oteborg, Sweden in the paper {\it Short-Lived Isotopes of Lanthanum, Cerium and Praseodymium Studied by SISAK-Technique} \cite{Aro74}. A uranium target was irradiated with 14 MeV neutrons and after chemical separation the cerium isotopes are identified using a continuous solvent extraction method. In addition to the observation of lighter-mass cerium isotopes ``Indirect evidence for the isolation of $\ge$ 5 sec $^{149}$Ce and 5-10 sec $^{150}$Ce is also presented.'' The adapted half-life of $^{149}$Ce is 5.3(2)s.

\subsection*{$^{150}$Ce}\vspace{-0.85cm}
The paper {\it Ground-State Bands in Neutron-Rich Even Te, Xe, Ba, Ce, Nd, and Sm Isotopes Produced in the Fission of $^{252}$Cf} published in 1970 reported the first identification of $^{150}$Ce by Wilhelmy {\it et al.} at Berkeley \cite{Wil70}. They measured $\gamma$-spectra following spontaneous fission of $^{252}$Cf and observed the first four $\gamma$-ray transitions of the $^{150}$Ce ground state band. They did not mention the first observation of $^{150}$Ce: ``The data, which in some of the cases can be correlated with previously reported results...'' implies that some of the cases were new observations.

\subsection*{$^{151}$Ce}\vspace{-0.85cm}
$^{151}$Ce was first identified in 1997 paper {\it Spectroscopy of neutron-rich odd-A Ce isotopes} by Hoellinger {\it et al.} in Strasbourg, France. Prompt $\gamma$-ray spectra from a $^{248}$Cm fission source were recorded with the EUROGAM II array. The paper only acknowledges the first observation of excited states in $^{151}$Ce: ``The level scheme of $^{151}$Ce, for which no excited states were known up to now,... '' They do not consider it a discovery of a new isotope because a lifetime measurement of $^{151}$Ce had been reported in a thesis from 1969 \cite{Wil69}.

\subsection*{$^{152}$Ce}\vspace{-0.85cm}
In 1990 Tago {\it et al.} reported the discovery of $^{152}$Ce at Kyoto, Japan: {\it Identification of $^{152}$Ce} \cite{Tag90}. A $^{235}$U target was irradiated in the Kyoto University Reactor. $^{152}$Ce was identified using the He-jet type on-line isotope separator (KUR-ISOL). ``From $\gamma$-ray spectra of mass separated fission products produced by KUR-ISOL, the Pr-K X-rays and two $\gamma$-rays are assigned to be generated by the $\beta$-decay of $^{152}$Ce.'' The measured half-life of 1.4(2) s is currently the accepted value. A previous claim of the observation of a $^{152}$Ce $\gamma$-ray transition \cite{Hil83} was later determined to be incorrect \cite{Kar87}.

\subsection*{$^{153-155}$Ce}\vspace{-0.85cm}
The isotopes $^{153}$Ce, $^{154}$Ce, and $^{155}$Ce were discovered at the Gesellschaft f\"ur Schwerionenforschung in Darmstadt, Germany in 1994 by M. Bernas {\it et al.} in the paper {\it Projectile fission at relativistic velocities: a novel and powerful source of neutron-rich isotopes well suited for in-flight isotopic separation.} \cite{Ber94}. A $^{235}$U beam of 750 MeV/nucleon bombarded a $^{208}$Pb target and the three new cerium isosopes was identified by the magnetic rigidity setting of the FRS fragment separator and a measurement of energy-loss and time-of-flight. A total of two, nine and four events for $^{153}$Ce, $^{154}$Ce, $^{155}$Ce, respectively. The identification of $^{153}$Ce and $^{155}$Ce were tentative due to the small number of events: ``Isotopes with less than 6 events registered are given in brackets.''

\section{Summary} \label{s:conclusion}
The discovery of the isotopes of cerium has been cataloged and the methods of their discovery discussed. The discovery of new isotopes is often linked to new technologies or techniques and is made based on limited statistics. Misidentifications can occur and it is important that the results are independently verified. For three of the cerium isotopes discussed, $^{130}$Ce, $^{139}$Ce and $^{145}$Ce the first reports of discovery were incorrect.

Finally, it is interesting to note that evidence for $^{128}$Ce, $^{129}$Ce, $^{150}$Ce and $^{151}$Ce reported in the refereed literature was based on $\gamma$-ray spectroscopy experiments where the authors did not acknowledge the first observation. They mentioned or must have assumed that the isotopes had already been discovered, most likely relying on conference proceedings or other unpublished reports.

\ack

This work was supported by the National Science Foundation under grant No. PHY-06-06007. JQG acknowledges the support of the Professorial Assistantship Program of the Honors College at Michigan State University.


\newpage

\section*{EXPLANATION OF TABLE}\label{sec.eot}
\addcontentsline{toc}{section}{EXPLANATION OF TABLE}

\renewcommand{\arraystretch}{1.0}

\begin{tabular*}{0.95\textwidth}{@{}@{\extracolsep{\fill}}lp{5.5in}@{}}
\textbf{TABLE I.}
	& \textbf{Discovery of Cerium Isotopes }\\
\\

Isotope &  Cerium isotope \\
Author & First author of refereed publication \\
Journal & Journal of publication \\
Ref. &  Reference  \\
Method & Production method used in the discovery: \\
    & FE: fusion evaporation \\
    & LPR: light-particle reactions (including neutrons) \\
    & MS: mass spectroscopy \\
    & RF: fission produced in reactors \\
    & SF: spontaneous fission \\
    & PR: projectile fragmentation or projectile fission \\
Laboratory &  Laboratory where the experiment was performed\\
Country &  Country of laboratory\\
Year & Year of discovery  \\
\end{tabular*}
\label{tableI}

\newpage
\datatables

\setlength{\LTleft}{0pt}
\setlength{\LTright}{0pt}


\setlength{\tabcolsep}{0.5\tabcolsep}

\renewcommand{\arraystretch}{1.0}


\begin{longtable}[c]{%
@{}@{\extracolsep{\fill}}r@{\hspace{5\tabcolsep}} llllllll@{}}
\caption[Discovery of Cerium Isotopes]%
{Discovery of cerium isotopes}\\[0pt]
\caption*{\small{See page \pageref{tableI} for Explanation of Tables}}\\
\hline
\\[100pt]
\multicolumn{8}{c}{\textit{This space intentionally left blank}}\\
\endfirsthead
Isotope & Author & Journal & Ref. & Method & Laboratory & Country & Year \\

$^{121}$Ce  & L. Zhankui & Phys. Rev. C &  Zha97   &  FE  &  Lanzhou  &  China  &  1997 \\
$^{122}$Ce  & J.F. Smith & Z. Phys. A &  Smi05   &  FE  &  Argonne  &  USA  &  2005 \\
$^{123}$Ce  & J.M. Nitschke & Nucl. Phys. A &  Nit84   &  FE  &  Berkeley &  USA  &  1984 \\
$^{124}$Ce  & D.D. Bogdanov & Nucl. Phys. A &  Bog78   &  FE  &  Dubna  &  Russia  &  1978 \\
$^{125}$Ce  & D.D. Bogdanov & Nucl. Phys. A &  Bog78   &  FE  &  Dubna  &  Russia  &  1978 \\
$^{126}$Ce  & D.D. Bogdanov & Nucl. Phys. A &  Bog78   &  FE  &  Dubna  &  Russia  &  1978 \\
$^{127}$Ce  & D.D. Bogdanov & Nucl. Phys. A &  Bog78   &  FE  &  Dubna  &  Russia  &  1978 \\
$^{128}$Ce  & D. Ward & Nucl. Phys. A &  War68   &  FE  &  Berkeley  &  USA  &  1968 \\
$^{129}$Ce  & J. Gizon & Nucl. Phys. A & Giz77   &  FE  &  Berkeley  &  USA  &  1977 \\
$^{130}$Ce  & G. Gerschel & Nuovo Cimento &  Ger65   &  LPR &  Orsay  &  France  &  1965 \\
$^{131}$Ce  & A. Norris & Nucl. Phys. &  Nor66   &  LPR &  Brookhaven  &  USA  &  1966 \\
$^{132}$Ce  & W.R. Ware & Phys. Rev. &  War60   &  LPR &  Rochester  &  USA  &  1960 \\
$^{133}$Ce  & B.J. Stover & Phys. Rev. &  Sto51   &  LPR &  Berkeley  &  USA  &  1951 \\
$^{134}$Ce  & B.J. Stover & Phys. Rev. &  Sto51   &  LPR &  Berkeley  &  USA  &  1951 \\
$^{135}$Ce  & J.B. Chubbuck & Phys. Rev. &  Chu48   &  LPR &  Berkeley  &  USA  &  1948 \\
$^{136}$Ce  & A.J. Dempster & Phys. Rev. &  Dem36   &  MS  &  Chicago  &  USA  &  1936 \\
$^{137}$Ce  & J.B. Chubbuck & Phys. Rev. &  Chu48   &  LPR &  Berkeley  &  USA  &  1948 \\
$^{138}$Ce  & A.J. Dempster & Phys. Rev. &  Dem36   &  MS  &  Chicago  &  USA  &  1936 \\
$^{139}$Ce  & M.L. Pool & Phys. Rev. &  Poo48   &  LPR &  Ohio State  &  USA  &  1948 \\
$^{140}$Ce  & F.W. Aston & Phil. Mag. &  Ast25   &  MS  &  Cambridge  &  England  &  1925 \\
$^{141}$Ce  & M.L. Pool & Phys. Rev. &  Poo48   &  LPR &  Ohio State  &  USA  &  1948 \\
$^{142}$Ce  & F.W. Aston & Phil. Mag. &  Ast25   &  MS  &  Cambridge  &  England  &  1925 \\
$^{143}$Ce  & M.L. Pool & Phys. Rev. &  Poo48   &  LPR &  Ohio State  &  USA  &  1948 \\
$^{144}$Ce  & W.H. Burgus & Nat. Nucl. Ener. Ser. &  Bur45   &  RF  &  Argonne  &  USA  &  1945 \\
$^{145}$Ce  & S.S. Markowitz & Phys. Rev. &  Mar54   &  RF  &  Brookhaven  &  USA  &  1954 \\
$^{146}$Ce  & A.A. Caretto & Phys. Rev. &  Car53   &  RF  &  Brookhaven  &  USA  &  1953 \\
$^{147}$Ce  & D.C. Hoffman & J. Inorg. Nucl. Chem. &  Hof64   &  RF  &  Los Alamos  &  USA  &  1964 \\
$^{148}$Ce  & D.C. Hoffman & J. Inorg. Nucl. Chem. &  Hof64   &  RF  &  Los Alamos  &  USA  &  1964 \\
$^{149}$Ce  & P.O. Aronsson & J. Inorg. Nucl. Chem. &  Aro74   &  LPR  &  Chalmers U  &  Sweden  &  1974 \\
$^{150}$Ce  & J.B. Wilhelmy & Phys. Rev. Lett. &  Wil70   &  SF  &  Berkeley  &  USA  &  1970 \\
$^{151}$Ce  & F. Hoellinger & Phys. Rev. C &  Hoe97   &  SF  &  Strasbourg  &  France  &  1997 \\
$^{152}$Ce  & I. Tago & Z. Phys. A &  Tag90   &  RF  &  Kyoto  &  Japan  &  1990 \\
$^{153}$Ce  & M. Bernas & Phys. Lett. B &  Ber94   &  PF  &  GSI  &  Germany  &  1994 \\
$^{154}$Ce  & M. Bernas & Phys. Lett. B &  Ber94   &  PF  &  GSI  &  Germany  &  1994 \\
$^{155}$Ce  & M. Bernas & Phys. Lett. B &  Ber94   &  PF  &  GSI  &  Germany  &  1994 \\
\end{longtable}

\newpage


\normalsize

\begin{theDTbibliography}{1956He83}

\bibitem[Zha97]{Zha97t} L. Zhankui, X. Shuwei, X.. Yuanxiang, M. Ruichang, G. Yuanxiu, G.W. Chunfang, H. Wenxue, and Z. Tianmei, Phys. Rev. C {\bf 56}, 1157 (1997)
\bibitem[Smi05]{Smi05t} J.F. Smith {\it et al.}, Phys. Lett. B {\bf 625}, 203 (2005)
\bibitem[Nit84]{Nit84t} J.M. Nitschke, P.A. Wilmarth, P.K. Lemmertz, W.D. Zeitz, and J.A. Honkanen, Z. Phys. A {\bf 316}, 249 (1984)
\bibitem[Bog78]{Bog78t} D.D. Bogdanov, A.V. Demyanov, V.A. Karnaukhov, M. Nowicki, L.A. Petrov, J. Voboril, and A. Plochocki, Nucl. Phys. A {\bf 307}, 421 (1978)
\bibitem[War68]{War68t} D. Ward, R.M. Diamond, and F.S. Stephens, Nucl. Phys. A {\bf 117}, 309 (1968)
\bibitem[Giz77]{Giz77t} J. Gizon A. Gizon, R.M. Diamond and F.S. Stephens, Nucl. Phys. A {\bf 290}, 272 (1977)
\bibitem[Ger65]{Ger65t} G. Gerschel, M. Pautrat, R.A. Ricci, J. Teillac, and J. Van Horenbeeck, Nuovo Cimento {\bf 37}, 1756 (1965)
\bibitem[Nor68]{Nor66t} A.E. Norris, G. Friedlander, and E.M. Franze, Nucl. Phys. {\bf 86}, 102 (1966)
\bibitem[War60]{War60t} W.R. Ware and E.O. Wiig, Phys. Rev. {\bf 117}, 191 (1960)
\bibitem[Sto51]{Sto51t} B.J. Stover, Phys. Rev. {\bf 81}, 8 (1951)
\bibitem[Chu48]{Chu48t} J.B. Chubbuck and I. Perlman, Phys. Rev. {\bf 74}, 982 (1948)
\bibitem[Dem36]{Dem36t} A. J. Dempster, Phys. Rev. {\bf 49}, 947 (1936)
\bibitem[Poo48]{Poo48t} M.L. Pool and N.L. Krisberg, Phys. Rev. {\bf 73}, 1035 (1948)
\bibitem[Ast25]{Ast25t} F.W. Aston, Phil. Mag. {\bf 49}, 1191 (1925)
\bibitem[Bur45]{Bur45t} W.H. Burgus, L. Weinberg, J.A. Seiler, and W. Rubinson, {\it Radiochemical Studies: The Fission Products}, Paper 184, p. 1195, National Nuclear Energy Series IV, 9, (McGraw-Hill, New York 1951)
\bibitem[Mar54]{Mar54t} S.S. Markowitz, W. Bernstein, and S Katcoff, Phys. Rev. {\bf 93}, 178 (1954)
\bibitem[Car53]{Car53t} A.A. Caretto, Jr. and S. Katcoff, Phys. Rev. {\bf 89}, 1267 (1953)
\bibitem[Hof64]{Hof64t} D.C. Hoffman and W.R. Daniels, J. Inorg. Nucl. Chem. {\bf 26}, 1769 (1964)
\bibitem[Aro74]{Aro74t} P.O. Aronsson, G Skarnemark, and M. Skarestad, J. Inorg. Nucl. Chem. {\bf 36}, 1689 (1974)
\bibitem[Wil70]{Wil70t} J.B. Wilhelmy, S.G. Thompson, R.C. Jared, and E. Cheifetz, Phys. Rev. Lett. {\bf25}, 1122 (1970)
\bibitem[Hoe97]{Hoe97t} F. Hoellinger {\it et al.}, Phys. Rev. C {\bf 56}, 1296(1997)
\bibitem[Tag90]{Tag90t} I. Tago, Y. Kawase, and K. Okano, Z. Phys. A {\bf 335}, 477 (1990)
\bibitem[Ber94]{Ber94t} M. Bernas {\it et al.}, Phys. Lett. B {\bf 331}, 19 (1994)

\end{theDTbibliography}

\end{document}